\documentclass[a4paper]{jpconf}
\usepackage{graphicx}
\begin{document}
\title{Transport through a double quantum dot with interdot repulsion}

\author{Yunori Nisikawa and Akira Oguri}

\address{Department of Material Science, Osaka City University,
 3-3-138 sumiyoshi-ku Osaka, Japan}

\ead{nisikawa@sci.osaka-cu.ac.jp}

\begin{abstract}
We study transport through a double quantum dot with 
interdot hopping $t$, intradot repulsion $U$ and interdot
 repulsion $U^{\prime}$, using the numerical renormalization group (NRG) method.
At half-filling, the conductances in two-terminal series and four-terminal parallel configuration are
 calculated via two phase shifts for quasi-particles of double quantum dot
connected to two noninteracting leads with hybridization strength $\Gamma$.
For small values of $t/\Gamma$ and $U^{\prime}/U$, 
 conductance in the two-terminal series configuration 
is suppressed to almost zero.
In this region, plateau of conductance in the four-terminal parallel configuration
appears and almost reaches a unitary limit value  $4e^{2}/h$ of two conducting modes.
For large values of $t/\Gamma$ or $U^{\prime}/U$, 
both conductances are suppressed to almost zero.
The conductance in the two-terminal series configuration almost reaches a
 unitary limit value $2e^{2}/h$ only around cross-over regions of electron-configuration in double quantum dot.
Through the behavior of the local charge and some thermodynamic quantities,
we discuss the relation between transport and electron-configuration.
\end{abstract}

A half-filled double quantum dot with interdot repulsion 
has been theoretically studied \cite{GLKprl,GLK,MRR} and some interesting phenomena, for example, 
quantum phase transition at a critical interdot repulsion
\cite{GLKprl}, have been predicted.
In this paper, focusing on two kind of conductance 
through a half-filled double quantum dot with interdot repulsion, 
we discuss the relation between transport and electron-configuration of double quantum dot  
through the behavior of the local charge and some thermodynamic
quantities, on the basis of numerical renormalization group (NRG) calculation.

\begin{figure}[ht]
\begin{center}
\includegraphics[width=10cm,height=4cm]{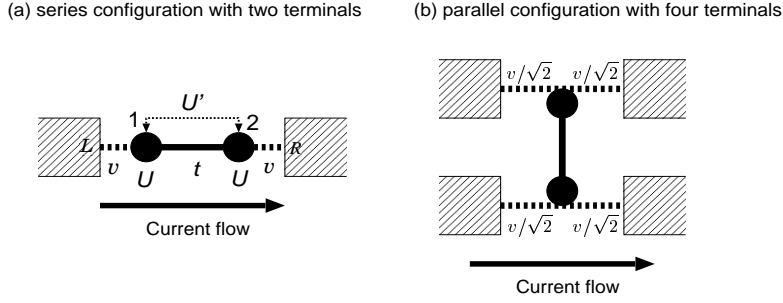}
\end{center}
\caption{\label{fig1} Schematic picture of (a) series configuration with
 two terminals and (b) parallel configuration with four terminals}
\end{figure}

We consider two-site Hubbard model with interdot repulsion,
which is connected to two non-interacting leads
at the left(L)  and right(R) by the symmetrical tunneling matrix elements 
$v$, as illustrated in Fig.\ \ref{fig1} (a).

The Hamiltonian is given by
$H=H_{D}+H_{\rm mix}+H_{\rm lead}$ with
\begin{eqnarray}
H_{D}&=&
\epsilon_d
\sum_{i=1}^{2}\sum_{\sigma} \,
 d^{\dagger}_{i\sigma}d^{\phantom{\dagger}}_{i\sigma}
-t \sum_{\sigma}
 \left(\,
 d^{\dagger}_{1\sigma}d^{\phantom{\dagger}}_{2\sigma}
 \,+\,
 d^{\dagger}_{2\sigma}d^{\phantom{\dagger}}_{1\sigma}
\right) +U\sum_{i=1}^{2}d_{i\uparrow}^{\dagger}d^{\phantom{\dagger}}_{i\uparrow}d_{i\downarrow}^{\dagger}d^{\phantom{\dagger}}_{i\downarrow}\nonumber\\
& &\mbox{}+U^{\prime}\sum_{\sigma\sigma^{\prime}}d_{1\sigma}^{\dagger}d^{\phantom{\dagger}}_{1\sigma}d_{2\sigma^{\prime}}^{\dagger}d^{\phantom{\dagger}}_{2\sigma^{\prime}}
\; ,\\
H_{\rm mix}&=&
v
 \sum_{\sigma}
 \left(\,
  d^{\dagger}_{1\sigma} \psi^{\phantom{\dagger}}_{L \sigma}
\,+\,
\psi^{\dagger}_{L \sigma}   d^{\phantom{\dagger}}_{1\sigma}
\,\right)
+ v
 \sum_{\sigma}
\left(\,
\psi^{\dagger}_{R\sigma} d^{\phantom{\dagger}}_{2\sigma}
      \,+\,
d^{\dagger}_{2\sigma} \psi^{\phantom{\dagger}}_{R\sigma}
      \,\right)  ,\\
H_{\rm lead}&=&
\sum_{\nu=L,R}
 \sum_{k\sigma}
  \epsilon_{k \nu}^{\phantom{0}}\,
         c^{\dagger}_{k \nu \sigma}
         c^{\phantom{\dagger}}_{k \nu \sigma}
\,,
\end{eqnarray}

where $d_{i\sigma}$ annihilates an electron with spin
$\sigma$ at site $i$ in the double quantum dot,
which is characterized by the interdot hopping
matrix element $t$, onsite energy $\epsilon_d$, 
intradot repulsion $U$ and interdot repulsion $U^{\prime}$.
In the lead at $\nu$ ($= L,\, R$), the operator
$c_{k \nu \sigma}^{\dagger}$ creates an electron
with energy $\epsilon_{k\nu}$ corresponding to
an one-particle state $\phi_{k\nu} (r)$.
The linear combinations of the conduction electrons
$\psi_{L \sigma}^{\phantom{\dagger}}$ and
$\psi_{R \sigma}^{\phantom{\dagger}}$  mixed with the electrons
 in dots labeled by $i=1$ and $2$, respectively,
where  $\psi_{\nu \sigma}^{\phantom{\dagger}}
= \sum_k c_{k \nu \sigma}^{\phantom{\dagger}}
\, \phi_{k\nu} (r_{\nu})$ and
$r_{\nu}$ is the position at the interface
in the lead $\nu$.
We assume that the hybridization strength
$\Gamma \equiv \pi v^2 \sum_k \left|\phi_{k\nu} (r_{\nu})\right|^2
\delta (\omega - \epsilon_{k \nu}^{\phantom{0}})$
is a constant independent of the frequency $\omega$ and $\nu$,
and take the Fermi energy $\mu$ to be $\mu=0$.

Our system has an inversion symmetry, so we can introduce
the even and odd-parity orbitals
as follows;
\begin{equation}
a_{\rm \sigma}=\frac{d_{1\sigma}+d_{2\sigma}}{\sqrt{2}}\; ,\ \
b_{\rm \sigma}=\frac{d_{1\sigma}-d_{2\sigma}}{\sqrt{2}}.\label{eo-o}
\end{equation}
Two phase shifts $\delta_{\rm e}$ and $\delta_{\rm o}$
of the quasi-particles with even and odd parities  
 characterize a local Fermi-liquid behavior of the whole system
 described by $H$ at low temperature.
We can obtain
 $\delta_{\rm e}$ and $\delta_{\rm o}$ from the
fixed-point eigen values of  NRG.
From two phase shifts defined with respect to the system described by
$H$, 
we can deduce not only 
the conductance $g_{\rm s}$ in the two-terminal series configuration
illustrated in Fig.\ \ref{fig1} (a) but also  
the conductance $g_{\rm p}$ in four-terminal parallel configuration illustrated in
Fig.\ \ref{fig1} (b) at $T=0$ as follows;
\begin{equation}
g_{\rm s}=\frac{2e^{2}}{h}\sin^{2}(\delta_{\rm e}- \delta_{\rm o})\; ,\ \
g_{\rm p}=\frac{2e^{2}}{h}\left(\sin^{2}\delta_{\rm e}+\sin^{2}\delta_{\rm o}\right).
\end{equation}
%

\begin{figure}[h]
\begin{minipage}{20pc}
\includegraphics[width=20pc]{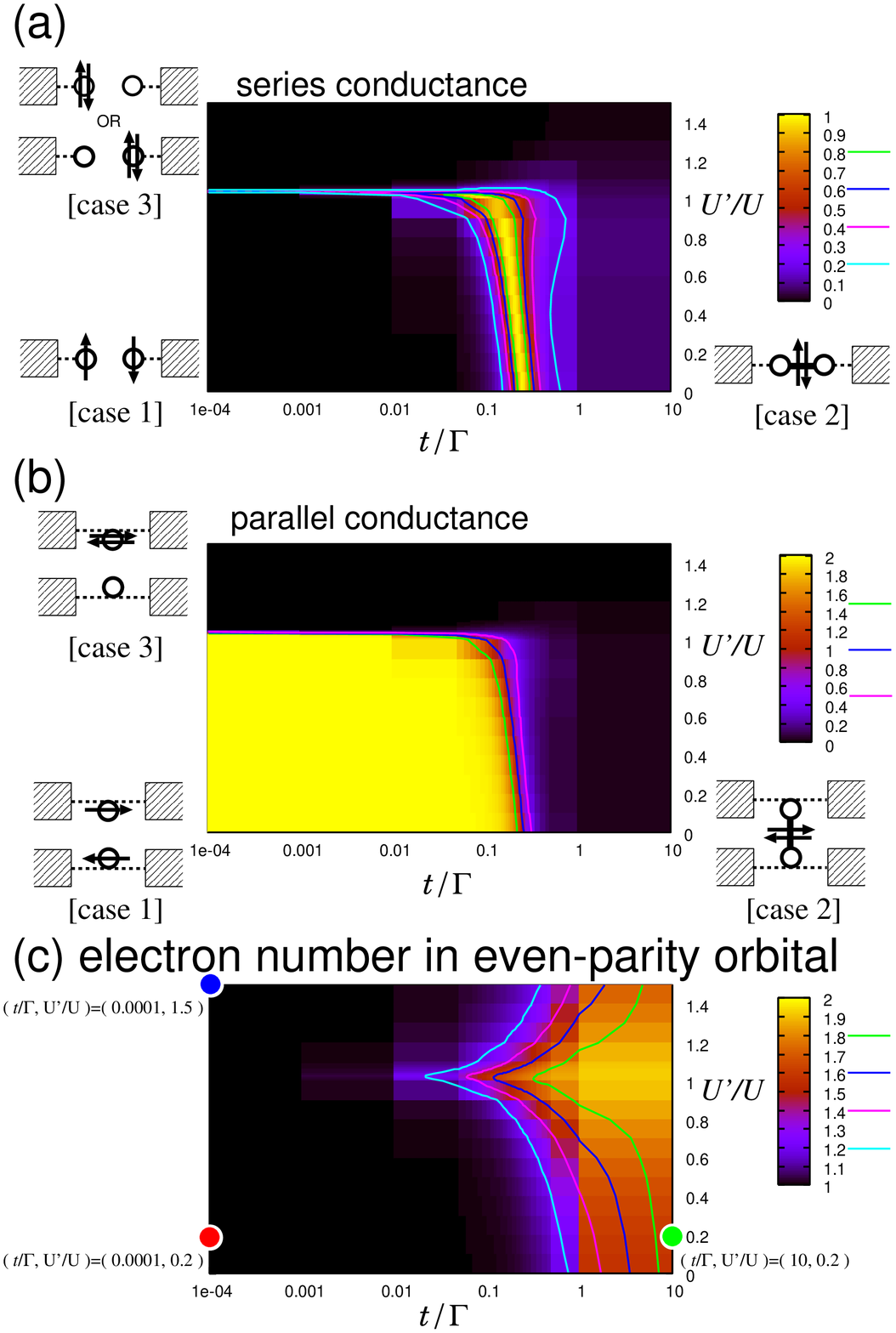}
\caption{\label{fig2}(a) series conductance $g_{s}/(2e^{2}/h)$, (b)
 parallel conductance $g_{p}/(2e^{2}/h)$  and
 (c) electron number in even-parity orbital plotted as functions
 $t/\Gamma$ and $U^{\prime}/U$. In (a) and (b),  schematic picture of
 electron-configuration for  case 1, case 2 and case 3 are presented.}
\end{minipage}\hspace{2pc}%
\begin{minipage}{16pc}
\includegraphics[width=16pc]{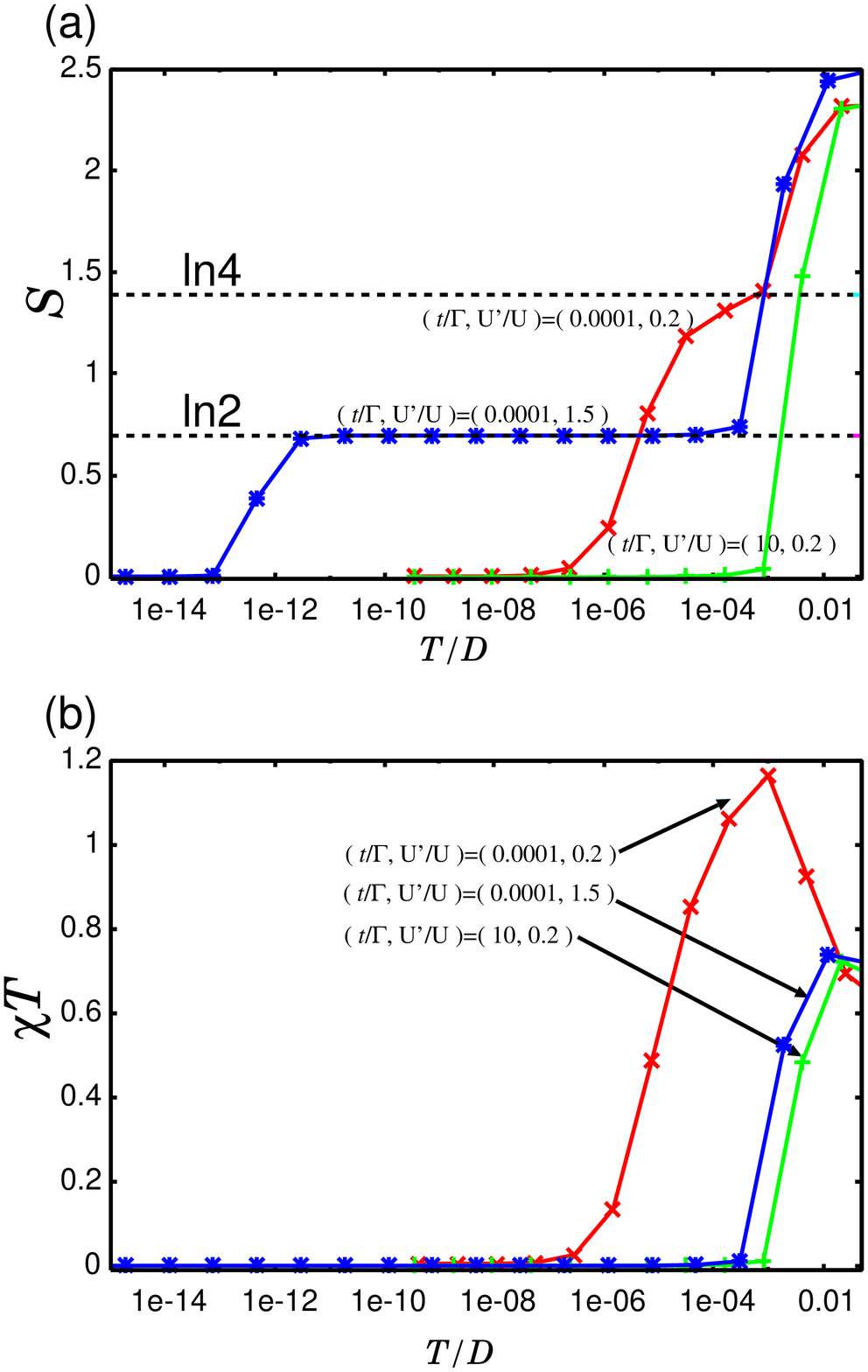}
\caption{\label{fig3}The temperature dependence of
(a) impurity entropy $S$ and (b) $\chi T$
for three points in $t/\Gamma - U^{\prime}/U$ plane shown in Fig.\ \ref{fig2}(c);$(t/\Gamma, U^{\prime}/U)=(0.0001,0.2)$, 
$(t/\Gamma, U^{\prime}/U)=(10,0.2)$ and 
$(t/\Gamma, U^{\prime}/U)=(0.0001,1.5)$.}
\end{minipage}
\end{figure}

We calculate series conductance  $g_{\rm s}$, parallel conductance
$g_{\rm p}$, electron number  
$n_{\rme}\equiv\sum_{\sigma}\langle a^{\dagger}_{\sigma}a_{\sigma}\rangle$ 
in even-parity orbital defined by Eq.(\ref{eo-o})
and some thermodynamic quantities 
at half-filling ($\epsilon_{d}=-U/2-U^{\prime}$)
for $10^{-4}\le t/\Gamma \le 10$ and $0\le U^{\prime}/U \le 1.5$.
In this paper, we fix $U/\Gamma$ to be $U/\Gamma=15$.
In Fig.\ \ref{fig2}, (a) $g_{\rm s}$, (b) $g_{\rm p}$ and (c) $n_{\rm e}$ 
are plotted as functions of $t/\Gamma$ and $U^{\prime}/U$.
First of all, to make discussion clear, we consider three limit case; 
case 1: $t/\Gamma \rightarrow 0$ and $U^{\prime}/U \rightarrow 0$
,
case 2: $t/\Gamma \rightarrow \infty$ and $U^{\prime}/U \rightarrow 0$
, and 
case 3: $t/\Gamma \rightarrow 0$ and $U^{\prime}/U \rightarrow \infty$.
In Fig.\ \ref{fig2}(a),  
we illustrate schematic picture of electron-configuration in double
quantum dot for these three limit case.
In the case 1, because of no interaction between two dots, 
our system is decoupled to two same systems 
composed of a single dot connected to a single lead.
At half-filling, each single dot is occupied by one electron to avoid intradot repulsion
$U$, as shown in the picture [case 1] in Fig.\ \ref{fig2}(a).
Schematic picture [case 2] in Fig.\ \ref{fig2}(a) represents that  
two electrons occupy the even-parity orbital stabilized by 
large interdot hopping $t$ in the case 2.
So, we can expect that the electronic states of the $n_{\rm e}\simeq 2$ 
region in Fig.\ \ref{fig2}(c) are similar to the state of the case 2.
The case 3 have been studied in detail by Galpin {\it et al.} using NRG\cite{GLKprl,GLK}. 
Either of two dot is occupied by two electrons to avoid strong
interdot repulsion $U^{\prime}$ 
which is much larger than intradot repulsion $U$, 
as shown in the picture [case 3] in Fig.\ \ref{fig2}(a).
Therefore, a degenerate configuration of two electrons
results in non-Fermi liquid ground state with ${\rm ln}2$ residual entropy.
From Fig.\ \ref{fig2}(a)
and above discussion, we find that 
the series conductance $g_{\rm s}$ almost reaches a unitary limit value
 $2e^{2}/h$ only around cross-over regions of 
electron-configuration in double quantum dot.
Next, we discuss the parallel conductance $g_{\rm p}$ in three limit cases. 
From the similar discussion for two-terminal series configuration, 
we can easily understand the schematic pictures of electron-configuration in four-terminal
parallel configuration for three limit case, shown in Fig.\ \ref{fig2}(b).
For the case 1, our system is decoupled to two same systems composed of single-dot
connected to two leads.
In each system, one electron occupies single dot as shown in the picture [case 1]
in Fig.\ \ref{fig2}(b), and causes Kondo effect. 
Therefore, the parallel conductance $g_{\rm p}$ 
reaches a unitary limit value equal to $4e^{2}/h$ of two conducting modes.
From 
Fig.\ \ref{fig2}(b), 
we indeed find that there is plateau of the parallel conductance 
$g_{\rm p}$, which almost reaches a unitary limit value 
$4e^{2}/h$,
for small values of $t/\Gamma$ and $U^{\prime}/U$.
The even-parity bonding orbital of double-dot is fully occupied by the
two electrons in the case 2.
In the case 3, two electrons fully occupy either of two dots
and no electron occupies the other, as shown in the picture [case 3] in
Fig.\ \ref{fig2}(b).
Therefore, the parallel conductance in both case 2 and 3 is 0.
This corresponds to the fact that  
the parallel conductance for large values of $t/\Gamma$ or $U^{\prime}/U$
is suppressed 
to almost zero, as shown in Fig.\ \ref{fig2}(b).
Fig.\ \ref{fig2}(a) is consistent with the phase diagram obtained by Mravlje {\it et al.}
using Gunnarsson and Sch\"{o}nhammer projection-operator method \cite{MRR}, apart from details.

We consider also impurity entropy $S$ and impurity spin
susceptibility $\chi$ 
calculated using NRG eigen-energy.
In Fig.\ \ref{fig3}, 
we show the temperature dependence of 
(a) impurity entropy $S$ and (b) $\chi T$ 
for three points in $t/\Gamma - U^{\prime}/U$ plane shown in Fig.\ \ref{fig2}(c);
$(t/\Gamma, U^{\prime}/U)=(0.0001,0.2)$, 
$(t/\Gamma, U^{\prime}/U)=(10,0.2)$ and 
$(t/\Gamma, U^{\prime}/U)=(0.0001,1.5)$.
%
For $(t/\Gamma, U^{\prime}/U)=(0.0001,0.2)$, 
impurity entropy shows steps of ln4 
at $T/D \simeq 10^{-4}$
 and
vanishes as temperature decreases.
There is a corresponding shoulder in impurity susceptibility. 
It implies that the ln4 step originates from the degree of spin freedom.
We can understand these thermodynamic behavior 
from consideration of the case 1.
Each $1/2$ -\ spin in single dot of two same systems for the case 1
is screened by Kondo effect in the same manner.
Therefore total entropy 2$\times$ln2=ln4 decays to 0 all at once.
The impurity entropy and susceptibility for 
$(t/\Gamma, U^{\prime}/U)=(10,0.2)$
decrease and vanish at high-temperature $T/D \simeq 10^{-3}$.
This is because, in the case 2 
assigned for $(t/\Gamma, U^{\prime}/U)=(10,0.2)$, 
the even-parity bonding orbital of double-dot 
stabilized by large hopping $t$ 
is fully occupied by the
two electrons and the freedom of spin and orbital are quenched at $T/D \simeq 10^{-3}$.
For $(t/\Gamma, U^{\prime}/U)=(0.0001,1.5)$, a plateau of ln2
is seen in the impurity entropy as temperature decreases.
The impurity spin susceptibility dose not change 
around temperature region $10^{-13}\le T/D \le 10^{-11}$ where the impurity entropy begins to drop
from ln2 to 0. 
Therefore, we find that the ln2 entropy is made up of the degree of orbital freedom.
The ln2 plateau corresponds to 
non-Fermi liquid ground state with  residual entropy in the case 3.
But for $(t/\Gamma, U^{\prime}/U)=(0.0001,1.5)$, a finite $t$  results in Fermi liquid
ground state and thereby the impurity entropy vanishes as temperature is lowered.

In summary, we have studied transport through a half-filled double quantum-dot with interdot repulsion
in two-terminal series and four-terminal parallel configuration, the 
electron number in even-parity orbital, impurity entropy and spin susceptibility, using NRG method.
Through the behavior of these quantities, 
we have presented the relation between transport and
electron-configuration of double quantum dot in wide parameter region.

\ack
We would like to thank V.~Meden for valuable discussions.
This work was supported
by JSPS Grant-in-Aid for Scientific Research (C).
Numerical computation was partly carried out
in Yukawa Institute Computer Facility.



\end{document}